\documentclass[twocolumn,aps,prd,showpacs,preprintnumbers,nofootinbib]{revtex4}

\usepackage[dvips]{graphicx}
\usepackage{bm,latexsym,amsmath,amssymb,amsfonts}

\begin{document}

\title{Spherical collapse of inhomogeneous dust cloud 
in the Lovelock theory}

\author{Seiju Ohashi\footnote{Email:ohashi@tap.scphys.kyoto-u.ac.jp}, 
Tetsuya Shiromizu\footnote{Email:shiromizu@tap.scphys.kyoto-u.ac.jp 
}} \affiliation{Department of Physics, Kyoto University, Kyoto 
606-8502, Japan}
\author{Sanjay Jhingan \footnote{Email:sanjay.jhingan@gmail.com}} 
\affiliation{Centre for Theoretical Physics, Jamia Millia Islamia, 
New Delhi 110025, India} 

\begin{abstract} 
We study gravitational collapse of a spherically symmetric inhomogeneous 
dust cloud in the Lovelock theory without cosmological constant. We show that 
the final fate of gravitational collapse in this theory depends on the 
spacetime dimensions. In odd dimensions the 
naked singularities formed are found to be massive. In the even dimensions, 
on the other hand, the naked singularities are found to be massless.
We also show that the curvature strength of naked singularity is independent 
of the spacetime dimensions in odd dimensions. However, it depends on the 
spacetime dimensions in even dimension. 
\end{abstract}
\maketitle

\section{Introduction}

Whether gravitational collapse results in a black hole or a naked
singularity is one of the crucial issues in classical general relativity. A singularity is not a ``{\it point}" in the spacetime and information from
such a place is in fact an end of causality in the theory. In order
to avoid such a breakdown of physics, it is often believed
that naked singularities cannot be formed under physically
reasonable conditions in classical general relativity (GR). This is known as the cosmic censorship
conjecture (CCC) \cite{Penrose:1969pc}. Broadly there are two
versions of the CCC, i.e., a strong cosmic censorship conjecture
(SCCC) and a weak cosmic censorship conjecture (WCCC). The SCCC
prohibits the formation of locally naked singularity, i.e.,
singularities are not visible to even nearby observers. On the other
hand, the WCCC prohibits the formation of globally naked singularity, allowing
a local observer moving around singularity to see it. Though there have
been many efforts to prove the CCC, it remains as one of the important unsolved problem in GR. 
Meanwhile, a considerable number of counterexamples have been
found  violating both the WCCC and SCCC.

In 1939 Oppenheimer and Snyder studied the gravitational
collapse of a spherically symmetric homogeneous dust cloud and found
that the collapse ends in a black hole \cite{Oppenheimer:1939ue}.
However,  in absence of a general proof of the CCC there
have been studies on collapse using various matter models like dust, 
radiation, scalar fields, and so on. In these exact solutions of Einstein
equations several counterexamples to the CCC have been found. For example, naked
singularities form in the collapse of spherically symmetric inhomogeneous 
dust models \cite{Yodzis:1973, Eardley:1978tr, Christodoulou:1984mz, Newman:1985gt, Joshi:1993zg,
Singh:1994tb, Jhingan:1996jb} (see Ref. \cite{Joshi:2008zz} and the
references therein). In general there are two types of singularities in 
spherical collapse, namely, shell-crossing and shell-focusing singularities. But, in what follows, 
we will focus only on the central shell-focusing singularities.

Recently, higher-dimensional spacetimes have generated considerable interest
in theoretical physics. These higher-dimensional models are motivated by string/M
theory. The case of collapse
of a inhomogeneous dust cloud in the simple extension of GR to higher dimensions
was well studied. It was shown there that the
singularity cannot be naked when the spacetime dimension is more
than six for smooth initial data \cite{ Ghosh:2001fb,
Goswami:2004gy, Goswami:2006ph} (see also Ref. \cite{Joshi:2008zz}
and the references therein). More precisely, one could confirm that
the SCCC holds in dimensions higher than six in such a naive extension of GR to higher dimensions. Clearly, the spacetime
dimensions play an important role in the final fate of the gravitational
collapse. However, it is expected that spacetime should be extremely
curved near singularities putting in doubt the validity of classical GR
in these regions. Indeed the string/M
theory predicts the higher curvature corrections to the Einstein
equation in its low-energy limit. In view of the above facts  it is important to study
gravitational collapse in such a theory. Though the CCC was originally formulated for
general relativity, it is worthwhile to investigate the implications of  stringy effects on it.
In this paper, we shall address this issue while focusing on the gravitational
collapse of spherical symmetric dust clouds. But there is no canonical
recipe to include higher-order curvature corrections to GR. Although string theory predicts such
corrections, we do not have a clear picture yet. In
this paper we shall adopt a special combination of higher
curvature corrections so that the field equations do not include the
higher derivative terms of the metric than the third order, that is, the Lovelock theory
\cite{Lovelock:1971yv}. There is no strong reason why the CCC should hold in 
the Lovelock theory, since energy conditions are generally violated in this theory. 
Nevertheless, if singularities do form in this theory,
it is important to know their features.

In the Lovelock theory the static black hole solutions of
a spherically symmetric vacuum were found in Refs.
\cite{Wheeler:1985qd,Cai:2003kt, Cai:2001dz}.
In Ref. \cite{Takahashi:2010ye,Takahashi:2010gz} the stability of a static vacuum black hole
solution was discussed. The gravitational collapse of a spherical homogeneous
dust cloud was analyzed in the dimensionally
continued gravity  \cite{Ilha:1996tc,Ilha:1999yn}(the restricted class of the Lovelock theory
\cite{Banados:1993ur}). It was shown there that the singularity will be covered
by event horizon as in the Oppenheimer-Snyder's case \cite{Oppenheimer:1939ue}.
Subsequently, the gravitational spherical collapse of inhomogeneous dust
cloud in the Gauss-Bonnet theory of gravity was addressed in Ref. 
\cite{Maeda:2006pm} and it was
shown that the final fate of collapse depends on the dimensions of spacetime. An exact solution
and complete analysis in the 5-D Gauss-Bonnet theory was given in Ref. \cite{JhinganGhosh}.
In five dimensions, singularities were found to be naked and massive.
From six to nine dimensions, singularities are naked and massless.
In dimensions higher than ten, singularities are always covered. Therefore, including the
Gauss-Bonnet corrections to GR makes the final state of 
collapse 
rather nontrivial. The gravitational collapse of null dust fluid was
also studied \cite{Maeda:2005ci,Nozawa:2005uy,Dehghani:2008yc}. 
These works also showed that the
Lovelock corrections modify the collapse scenario in a nontrivial way.

In this paper we investigate the gravitational 
collapse of a spherical inhomogeneous dust cloud in the Lovelock gravity, without
cosmological constant and in any dimensions. We show that
singularities do form in the final stage of gravitational
collapse in general and that they could be naked for some initial dates.
Comparing with the Gauss-Bonnet gravity, we found that the Lovelock
term changes the nature of singularities. In the Gauss-Bonnet case,
singularities which appear in more than six dimensions are  massless.
In the Lovelock theory with  odd dimensions, we see
that singularities are massive. On the other hand, singularities
are massless in the even dimensions. This is an unique feature of the Lovelock
gravity.

The rest of this paper is organized as follows. In Sec. II, we briefly review the Lovelock theory. In Sec. III, we derive the
basic equations for gravitational spherical dust collapse in the
Lovelock theory. In Sec. IV, we analyze the nature of
the singularity and apparent horizon.
In Sec. V, we show the existence of null geodesics from
the singularities.
In Sec. VI, we analyze the strength of the naked singularities.
Finally we will give our conclusion and discussion in Sec. VII. In the Appendix A, 
we discussed the junction between inner and outer solutions. In the Appendix B, 
the analysis of the homogeneous case will be briefly given. 
In this paper, we employ the units of $c=1$ and $8\pi G=1$, where $c$
is the speed of light and $G$ is the gravitational constant in 
higher dimensions.

\section{Lovelock theory}

In this section we give an overview of the Lovelock theory in 
$D=n + 2$-dimensional spacetimes.
This is the most general theory of gravity satisfying
following three conditions:
\begin{enumerate} 
\item The field equations are written in terms of a
symmetric rank-2 tensor. 
\item The theory is consistent with the conservation law
of the energy-momentum tensor. 
\item The theory does not include higher
than third order derivatives. 
\end{enumerate}

The Lagrangian of the theory is given by
\begin{equation}
\mathcal{L}=\sum_{m=1}^k\frac{a_m}{m}\mathcal{L}_m,
\end{equation}
%
where
%
\begin{equation}
\mathcal{L}_{m}=\frac{1}{2^m}\delta_{\mu_1 \mu_2 \dots \mu_{2m-1}
\mu_{2m}}^{\nu_1 \nu_2 \dots \nu_{2m-1} \nu_{2m}}R_{\nu_1
\nu_2}{}^{\mu_1 \mu_2 }\dots
R_{\nu_{2m-1}\nu_{2m}}{}^{\mu_{2m-1}\mu_{2m}}{}.
\end{equation}
%
In the above, $R_{\mu\nu}{}^{\rho\sigma}$ is the Riemann curvature tensor, and
$\delta_{\mu_1 \mu_2 \dots \mu_{2m-1}
\mu_{2m}}^{\nu_1 \nu_2 \dots \nu_{2m-1} \nu_{2m}}$ is the generalized and
 totally
antisymmetric Kronecker delta. $\lbrace a_m \rbrace $ are arbitrary constants
which cannot be determined by the theory itself. The suffixes $\mu_1 \dots \mu_{2m}$ and
$\nu_1 \dots \nu_{2m}$ run from $1$ to $D$, and  $k$ is a constant depending
on the spacetime dimensions, defined by $k=[(D-1)/2]$ ($[x]$
is the integer part of $x$). Throughout this paper we suppose that $a_m$ are positive.

The field equations for this Lagrangian are
%
\begin{eqnarray}
\mathcal{G}_{\mu}^{\nu} & = &
-\sum_{m=1}^{k}\frac{1}{2^{m+1}}\frac{a_m}{m}\delta_{\mu \mu_1 \mu_2
\dots \mu_{2m-1} \mu_{2m}}^{\nu \nu_1 \nu_2 \dots \nu_{2m-1} \nu_{2m}}
\nonumber \\
&& \times R_{\nu_1 \nu_2}{}^{\mu_1 \mu_2 }\dots
R_{\nu_{2m-1}\nu_{2m}}{}^{\mu_{2m-1}\mu_{2m}} \notag \nonumber \\ &=
& T_\mu^\nu,
\end{eqnarray}
%
where $T^\nu_\mu$ is energy-momentum tensor of matter. It should be remembered that
$\nabla^\mu \mathcal{G}_{\mu}^{\nu}=0$ holds, where
$\nabla_\mu$ is the covariant derivative with respect to the metric $g_{\mu\nu}$.

There is an exact solution of the vacuum, static and spherical symmetric
black hole \cite{Wheeler:1985qd,Cai:2003kt,Cai:2001dz}. The metric
is given by
%
\begin{equation}
ds^2=-f(r)dt^2+\frac{1}{f(r)}dr^2+r^2\gamma_{ij}dx^idx^j, \label{solution}
\end{equation}
%
where $\gamma_{ij}$ is the line element of $n$-dimensional constant
curvature hypersurface and $f(r)$ is
%
\begin{equation}
f(r)=\kappa -r^2\psi (r) \label{deff}.
\end{equation}
%
In the above, $\kappa$ is the constant curvature of
the $n$-dimensional hypersurfaces which take values $-1,0 \text{\ and\ } 1$.
In this paper we are interested only in asymptotically flat, spherically
symmetric spacetime and  then set $\kappa =1$.
The function $\psi (r)$ is a solution to the algebraic equation
%
\begin{equation}
W[ \psi ] =\sum_{m=2}^{k} \bigg[ \frac{a_m}{m}\bigg\{
\prod_{p=1}^{2m-2}(n-p)\bigg\} \psi^m \bigg] +\psi
=\frac{\mu}{r^{n+1}} .\label{defpsi}
\end{equation}
%
Here $\mu$ is a constant proportional to the ADM mass.

Under the assumption that all the coefficient $a_m$ are positive,
$W[\psi ]$ is a monotonically increasing function of $\psi$ when $\psi >0$.
Therefore, positive $\psi$ has a unique solution to Eq. (\ref{defpsi}).
When we take $r \to \infty$, the positive solution goes to zero as
%
\begin{equation}
\psi =\frac{\mu}{r^{n+1}}+\mathcal{O}(\frac{1}{r^{n+2}}) .
\end{equation}
%
From this fact and the metric form of Eq. (\ref{deff}), it is easy to see that
the solution corresponds to an asymptotically flat spacetime.
A negative $\psi$ could also be the solution. But in this case,
the solution corresponds to an asymptotically anti-deSitter (AdS) spacetime.
In this paper we shall consider  on  the positive $\psi$ solution because we are interested in 
asymptotically flat spacetimes.

The horizon is located at $f=0$. Then the equation for the horizon is
%
\begin{equation}
\psi =\frac{1}{r^2} \label{horizon1}
\end{equation}
%
and Eq. (\ref{defpsi}) implies 
%
\begin{equation}
W[ \psi ] =\mu \psi^{(n+1)/2} \label{horizon2}
\end{equation}
%
at the horizon. The solution to Eq. (\ref{horizon2}) depends on spacetime dimension,
$n$.
In even dimensions, Eq. (\ref{horizon2}) becomes
%
\begin{eqnarray}
 -\mu \psi^{(n-1)/2}+\sum_{m=2}^{n/2} \bigg[ \frac{a_m}{m}\bigg\{ \prod_{p=1}^{2m-2}(n-p)\bigg\} \psi^{m-1} \bigg] +1=0. \nonumber \\
\label{horizon3}
\end{eqnarray}
%
It is easy to see that there are always solutions to Eq. (\ref{horizon3}).
In odd dimensions, on the other hand, Eq. (\ref{horizon2}) becomes
%
\begin{eqnarray}
& & \Bigl( \frac{2a_{\frac{n+1}{2}}}{n+1}\bigg\{ \prod_{p=1}^{n-1}(n-p)\bigg\} -\mu \Bigr) \psi^{(n-1)/2}
\nonumber \\
& & +\sum_{m=2}^{(n-1)/2} \bigg[ \frac{a_m}{m}\bigg\{ \prod_{p=1}^{2m-2}(n-p)\bigg\} \psi^{m-1} \bigg] +1=0.
\label{horizon4}
\end{eqnarray}
%
There are the solutions to the equation above if
%
\begin{equation}
\frac{2a_{\frac{n+1}{2}}}{n+1}\bigg\{ \prod_{p=1}^{n-1}(n-p)\bigg\} -\mu  <0 
\end{equation}
%
is satisfied. In odd dimensions, there is a tendency that spacetimes with small mass
have no horizons. This is the characteristic feature which is different
from even-dimensional cases.
For example, in the five-dimensional asymptotically flat solution, $f(r)$ becomes
%
\begin{equation}
f(r)=1+\frac{r^2}{2a_2}\left( 1-\sqrt{1+\frac{4a_2\mu}{r^4}} \right) .
\end{equation}
%
The the event horizon is given by
%
\begin{equation}
r=\sqrt{\mu -a_2} .
\end{equation}
%
Since $a_2>0$, the size of the Gauss-Bonnet black hole is smaller than
that of the corresponding five-dimensional Schwarzschild black hole.
Therefore, the Gauss-Bonnet correction weaken the gravity.
Note that this solution has the naked singularity when $\mu \leq a_2$.

For our current study, the vacuum exact solution mentioned above stands for 
 the outer region of the collapsing dust cloud. Actually, we can join this outer solution with the inner cloud solution (see the
appendix A for the details).

\section{Spherical collapse of dust clouds}

In this section, we will consider the spherical collapse of dust clouds
in the Lovelock theory. We first write down the basic equations.

The energy-momentum tensor of dust cloud is
%
\begin{equation}
T_\nu^\mu=\varepsilon (t,r)u_\nu u^\mu ,
\end{equation}
%
where $\varepsilon (t,r)$ is the energy density, which is supposed to be
positive, and $u^{\mu}$ is the velocity. The metric is
%
\begin{equation}
ds^2=-B(t,r)^2dt^2+A(t,r)^2dr^2+R(t,r)^2d\Omega_n^2 \label{metric1}.
\end{equation}
%
Here $A(t,r)$, $B(t,r)$ and $R(t,r)$ are arbitrary functions of
$t$ and $r$, and $d\Omega_n^2$ is the line element of the $n$-dimensional unit
sphere. Using the gauge freedom of coordinate $r$, we can always set $R(0,r)=r$.
In the current paper, we assume $\partial_t R < 0$ at initial surface. 
This means that the dust cloud is initially collapsing.

From the conservation equation of the energy-momentum tensor
%
\begin{equation}
\nabla_{\nu}T^{\nu}_{\mu}=0,
\end{equation}
%
we have two equations
%
\begin{align}
 \varepsilon  \frac{B^{\prime}}{B} &=0  ,\label{conserve1}\\
 \dot{\varepsilon} +\varepsilon \left( \frac{\dot{A}}{A}+n\frac{\dot{R}}{R}\right) &=0  .\label{conserve2}
\end{align}
%
In the above the dot and prime denote
the partial derivative of $t$ and $r$, respectively.

It is easy to see that Eq. (\ref{conserve1}) implies $B=B(t)$.
Then, without a loss of generality, we can set $B=1$.
Now the metric becomes
%
\begin{equation}
ds^2=-dt^2+A(t,r)^2dr^2+R(t,r)^2d\Omega_n^2 \, ,\label{metric2}
\end{equation}
%
and the gravitational field equations for this simplified metric are
%
\begin{widetext}
\begin{eqnarray}
\sum_{m=1}^{k}a_m\prod_{p=0}^{2m-2}(n-p)\left(\left( \frac{\dot{R}}{R}\right)^2-\left( \frac{R^{\prime}}{RA}\right)^2  +\frac{1}{R^2}\right)^{m-1}
\left( \frac{\ddot{R}}{R} +\frac{\left( n-\left( 2m-1\right) \right)}{2m}\left( \left( \frac{\dot{R}}{R}\right)^2-\left( \frac{R^{\prime}}{RA}\right)^2  +\frac{1}{R^2}\right) \right)
                   =0, \label{LL1}
\end{eqnarray}
%
%
\begin{eqnarray}
& & \sum_{m=1}^{k}a_m\prod_{p=0}^{2m-2}(n-p)\left( \left( \frac{\dot{R}}{R}\right)^2-\left( \frac{R^{\prime}}{RA}\right)^2  +\frac{1}{R^2}\right)^{m-1}
\nonumber \\
& & \times \Biggl( -\frac{R^{\prime \prime }}{RA^2}+\frac{R^{\prime}A^{\prime}}{RA^3} +\frac{\dot{R}\dot{A}}{RA} +\frac{\left( n-\left( 2m-1\right) \right) }{2m} \Bigl( \left( \frac{\dot{R}}{R}\right)^2-\left( \frac{R^{\prime}}{RA}\right)^2  +\frac{1}{R^2}\Bigl) \Biggr)
                   =\varepsilon (t,r) ,\label{LL2}
\end{eqnarray}
%
and
%
\begin{eqnarray}
\sum_{m=1}^{k}a_m\prod_{p=0}^{2m-2}(n-p)\left( \left( \frac{\dot{R}}{R}\right)^2-\left( \frac{R^{\prime}}{RA}\right)^2  +\frac{1}{R^2}\right)^{m-1}  \left( \frac{\dot{R}^{\prime}}{RA^2}-\frac{R^{\prime}\dot{A}}{RA^3} \right)
                  =0 \label{LL3} .
\end{eqnarray}
\end{widetext}
%

From Eq. (\ref{LL3}), we see that
%
\begin{equation}
 \frac{\dot{R}^{\prime}}{RA^2}-\frac{R^{\prime}\dot{A}}{RA^3} =0 
\end{equation}
%
holds. This equation can be easily integrated as
%
\begin{equation}
A=\frac{R^{\prime}}{W(r)} \label{A},
\end{equation}
%
where $W(r)$ is the arbitrary function of $r$.
Introducing $L$ by
%
\begin{equation}
L\equiv \sum_{m=1}^k\frac{a_m}{2m}\prod_{p=0}^{2m-2}(n-p)\left( \dot{R}^2+1-W^2\right)^{m}\frac{1}{R^{2m}} \label{defL},
\end{equation}
%
and using Eq. (\ref{A}), we can rewrite Eqs. (\ref{LL1})
and (\ref{LL2}) in the following simple forms
%
\begin{align}
\frac{1}{\dot{R}R^n}\frac{\partial}{\partial t}\left( LR^{n+1}\right) &=0 \label{simpleLL1} ,\\
\frac{1}{R^{\prime}R^n}\frac{\partial}{\partial r}\left( LR^{n+1}\right) &=\varepsilon (t,r) \label{simpleLL2}.
\end{align}
%
Then the integration of the above equations implies
%
\begin{equation}
L=\frac{n}{2}\frac{F(r)}{R^{n+1}} \label{LL6},
\end{equation}
%
where $F(r)$ is a function of $r$, which can be interpreted as a
sort of mass function defined as
%
\begin{equation}
F(r)=\frac{2}{n}\int_0^r \varepsilon R^nR^{\prime} dr  .
\end{equation}
%
Indeed, $F(r)$ is proportional to the generalized Misner-Sharp mass 
\cite{Maeda:2011ii}. Using $F(r)$, Eq. (\ref{simpleLL2}) can be reexpressed as
%
\begin{equation}
\varepsilon =\frac{n}{2}\frac{F(r)^{\prime}}{R^nR^{\prime}}. \label{simpleLL3}
\end{equation}
%

For later convenience, we shall introduce the coefficients $c_l$ by
%
\begin{equation}
c_l =\begin{cases}
     1 & \text{if}\ \  l=1\\
     \frac{a_l}{l}\prod_{p=1}^{2l-2}(n-p) & \text{if}\ \  2\leq l \leq k .
     \end{cases}
\end{equation}
%
Using this, Eq. (\ref{LL6}) becomes
%
\begin{equation}
\sum_{m=1}^k c_m \left( \dot{R}^2+1-W^2\right)^{m}\frac{1}{R^{2m}}=\frac{F(r)}{R^{n+1}} \label{basiceq}.
\end{equation}
%
This is the key equation for the gravitational collapse.

From Eq. (\ref{simpleLL3}), we can see two types of singularities which correspond
 to  blowing up of energy density, that is,
the shell-focusing singularities which form at at $R=0$, and
the shell-crossing singularities at $R^{\prime}=0$. In this paper 
we assume $R^{\prime}>0$ to exclude the shell-crossing singularities
from our consideration. 

To make the initial surface regular at the center ($r=0$),
the function $F(r)$ should behave as
%
\begin{equation}
F(r)=r^{n+1}M(r) \label{regularity},
\end{equation}
%
where $M(r)$ is some regular function on initial surface.

For later convenience, we introduce the function $v(t,r)$ through
%
\begin{equation}
R(t,r)=rv(t,r) \label{v}.
\end{equation}
%
Then our current initial conditions of $R(t,r)$ are expressed as
%
\begin{align}
&v(0,r)=1 ,\\
&\dot{v}(0,r)< 0.
\end{align}
%
The singularities at $R=0$ are located at the orbit satisfying $v(t,r)=0$.

Before closing this section, we have one remark. At first glance, 
\begin{equation}
\sum_{m=1}^{k}a_m \prod_{p=0}^{2m-2}(n-p)\left( \left( \frac{\dot{R}}{R}\right)^2
-\left( \frac{R^{\prime}}{RA}\right)^2  +\frac{1}{R^2}\right)^{m-1} =0
\end{equation}
is also the solution to Eq. (\ref{LL3}).
Combining above with Eq. (\ref{LL1}), we have
\begin{eqnarray}
& & \sum_{m=1}^{k}a_m\prod_{p=0}^{2m-2}(n-p)\frac{\left( n-\left( 2m-1\right) \right)}{2m}
\nonumber \\
& &~~\times \left(\left( \frac{\dot{R}}{R}\right)^2
-\left( \frac{R^{\prime}}{RA}\right)^2  +\frac{1}{R^2}\right)^{m}=0.
\end{eqnarray}
Together with Eq. (\ref{LL2}), we can show that this implies $\varepsilon (t,r)=0$.
This means that the spacetimes must be vacuum, which is a solution 
but not our interest here.

\section{Apparent horizons and singularities}

In this section  we analyze the structure of spacetimes obtained in the previous section.
We shall focus on the behavior of trapped surfaces and apparent horizons surrounding the singularities.

We shall  investigate the even and odd-dimensional cases separately
since it is clear from Eqs. (\ref{horizon3}) and (\ref{horizon4}) that
they should have quite different features.
As stated in Sec. II, we assume that all coefficients $c_l$ are positive.
Here we also take $W(r)=1$, which corresponds to the ``marginally bound case".

First of all, it should be noted that bounce ($\dot{v}=0$ and $v \neq 0$) cannot occur in this case for initially collapsing configurations. 
Then we will be able to expect the appearance of the shell-focusing singularity in general. 
Using Eqs. (\ref{regularity}) and (\ref{v}), Eq. (\ref{basiceq}) becomes 
%
\begin{equation}
\sum_{m=1}^k c_m \left( \frac{\dot{v}}{v}\right)^{2m}=\frac{M(r)}{v^{n+1}}. \label{basiceq2}
\end{equation}
%
Substituting $\dot{v}=0$ into Eq. (\ref{basiceq2}), then we have
%
\begin{equation}
M(r)=0
\end{equation}
%
for the bounce. This contradicts our assumption that $\varepsilon$ is positive. 
From this fact, bounce cannot occur and the cloud continues to collapse. We can also see from this that the range of $v$ is from $0$ to $1$.

A trapped surface is described by 
the surfaces which satisfy $dR/dt|_{\pm} <0$, where $d/dt|_+$ and $d/dt|_-$ are 
derivatives along the outgoing and ingoing null geodesics respectively. For the current metric form, 
$dR/dt|_\pm$ becomes 
%
\begin{align}
\frac{dR}{dt}\bigg|_+ &=\dot{R}+\frac{dr}{dt}\bigg|_+ R'=\dot{R}+1  , \label{trap1}\\
\frac{dR}{dt}\bigg|_- &=\dot{R}+\frac{dr}{dt}\bigg|_- R'=\dot{R}-1, \label{trap2}
\end{align}
%
where we used the fact of $dr/dt|_\pm =\pm R'^{-1}$.
The apparent horizon is defined to be the boundary of trapped regions. In our case it is
easy to see that the location of the apparent horizons can be found through the solving
of
%
\begin{equation}
g^{\mu \nu}R_{,\mu}R_{,\nu}=0.
\end{equation}
%
For the metric of Eq. (\ref{metric2}), this condition becomes
%
\begin{equation}
{\dot{R}^2}=1. \label{conR2}
\end{equation}
%

\subsection{Even dimensions}

\begin{figure}[!htb]
\includegraphics[width=6.5cm,clip]{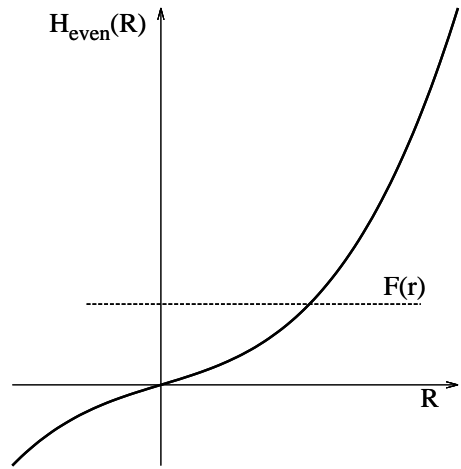}\\
\caption{The profile of function $H_{even}(R)$. The thick line is $H_{even}(R)$ and the dotted line is
$F(r)$. $H_{even}(R)$ is the odd function and we assume that all coefficients are positive.
Therefore, $H_{even}$ is monotonically increasing function for $R>0$ and  $H_{even}(0)=0$. }
\label{Grapheven}
\end{figure}

We first consider the apparent horizon in even dimensions.
Using Eq. (\ref{conR2}), Eq. (\ref{basiceq}) implies
%
\begin{equation}
H_{even}(R)\equiv \sum_{m=1}^{k=n/2} c_mR^{n-2m+1}=F(r)\,, \label{AHeven}
\end{equation}
%
which determines the location of the apparent horizons.
Note that the left-hand side of the equation above is an odd function of $R,$ and $F(r)$ is
a monotonically increasing function of $r,$ with $\lim_{r\to 0}F(r)=0$. We depict the
behavior of the left-hand side of Eq. (\ref{AHeven}) in Fig. \ref{Grapheven}.
As shown in the figure, for any  given finite $r$, we always have a solution to Eq. (\ref{AHeven}).
This implies that the apparent horizons always exist in these spacetimes
and all singularities which occur at
finite $r>0$ (but, $R=0$) will be the inside of the apparent horizon. Therefore, only the central singularity at $r=0$ could be
naked. In addition, they are massless because $F(0)=0$, in even dimensions. (Note that $F(r)$ is proportional to a quasilocal mass.)
This is the same property as in the Gauss-Bonnet case.
In the Gauss-Bonnet case, all singularities except for
$r=0$ were found to be covered by apparent horizons in any dimensions higher than six  \cite{Maeda:2006pm}.

Next, let us confirm that the region inside the apparent horizon is trapped. 
From Eq. (\ref{basiceq}), $\dot{R}^2$ continues to increase during the collapse, 
which means $R$ decreases. This implies $\dot{R}$ is monotonically decreasing 
because we assume that the dust cloud is initially collapsing. Then, we can see 
that the region inside of the apparent horizon($\dot R=-1$) is actually trapped, $dR/dt|_{\pm} < 0$, from 
Eqs. (\ref{trap1}) and (\ref{trap2}). 

\subsection{Odd dimensions}

\begin{figure}[!htb]
\includegraphics[width=6.5cm,clip]{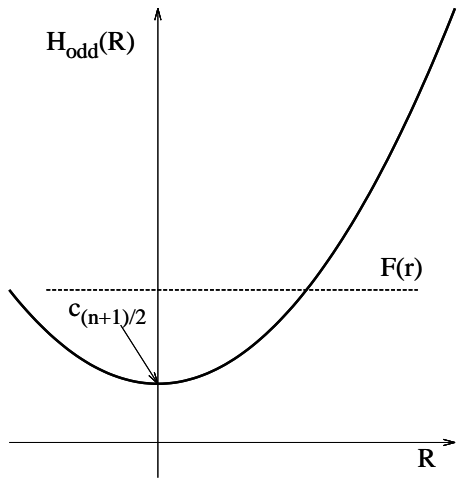}\\
\caption{The profile of the function $H_{odd}(R)$. The thick line is $H_{odd}(R)$ and
the dotted line is $F(r)$. $H_{odd}(R)$ is the even function and we assume that 
all coefficients are positive. Therefore  $H_{odd}$ is monotonically increasing function
of $R>0$ and $H_{odd}(0)=c_{(n+1)/2}>0$.
The figure shows that $r$ satisfying $F(r)<c_{(n+1)/2}$ have no solution to Eq. (\ref{AHodd}).}
\label{Graphodd}
\end{figure}

As in even dimensional cases, Eqs. (\ref{conR2}) and (\ref{basiceq}) give
%
\begin{equation}
H_{odd}(R)\equiv \sum_{m=1}^{k=(n+1)/2} c_mR^{n-2m+1}=F(r)\, , \label{AHodd}
\end{equation}
%
which determines the location of the apparent horizons.
Contrary to what we had in even dimensions, the left-hand side of Eq. (\ref{AHodd}) is the even function of $R$
(See Fig. \ref{Graphodd}).
In the range of $0\leq r \leq r_{ah}$, we see that there are no solutions,
where $r_{ah}$ is determined through
%
\begin{equation}
F(r_{ah})=c_{(n+1)/2} \label{rah}.
\end{equation}
%
Hence the singularities which occurs in the range $0\leq r \leq r_{ah}$  are not
wrapped by the apparent horizons. They could be naked and massive ( 
$F(r^{*})\geq 0$ where $0\leq r^{*}\leq r_{ah}$) in odd dimensions. 
This result is in agreement with what is observed in  the five-dimensional 
Gauss-Bonnet case \cite{Maeda:2006pm}. For the Gauss-Bonnet case in any dimensions higher than six, 
however, the singularities at finite $r$ shall always be safely wrapped. 
We also see that the region inside of the apparent horizon is trapped region 
from the same argument in even-dimensional cases.

\section{Null geodesics from singularities}

In this section, we shall analyze  whether singularities, which are not wrapped by apparent horizons,
can be naked or not. To this end we examine the existence of future-directed outgoing radial null
geodesics from these singularities. Again, we need to analyze odd- and even-dimensional cases separately.

From the metric of Eq. (\ref{metric2}), the outgoing radial null geodesic equation becomes
%
\begin{equation}
\frac{dt}{dr}=R^{\prime}. \label{nullgeodesiceq}
\end{equation}
%
To show the existence of solution for this differential equation from the singularity, we shall employ the fixed-point method
\cite{Christodoulou:1984mz, Newman:1985gt, Maeda:2006pm}. We define a new function
%
\begin{equation}
x\equiv \frac{R}{(r-r_s)^q},
\end{equation}
%
where $q$ is a positive constant and $r_s$ is the location of singularities.
Using $x$, the null geodesic equation can be rewritten as
%
\begin{equation}
\frac{dx}{dr}+\frac{qx}{r-r_s}=\Xi (x,r) ,\label{nullgeo}
\end{equation}
%
where $\Xi (x,r)$ is given by
%
\begin{equation}
\Xi (x,r)=\frac{R^{\prime}}{(r-r_s)^q}\left( 1+\dot{R} \right) .\label{deftheta}
\end{equation}
%
One may consider the cases in which
the function $\Xi (x,r)$ can be expanded around $r=r_s$
%
\begin{eqnarray}
\Xi (x,r) & \simeq &  \frac{g_0(q)x+\sum_{i=1}^{N}g_i(q)x^{\alpha_i}}{r-r_s}
+h(q,x)(r-r_s)^a \nonumber \\
& & ~~+o((r-r_s)^a), \label{expand}
\end{eqnarray}
%
where $a$  $(>-1)$ and $\alpha_i\neq 1$ are constants. $g_0,g_i$ and $h$ are some functions.
$N\geq 1$ is an integer.
We will be able to use the following useful lemma to show the existence of
radial null geodesic from singularities
\vspace{3mm}

{\it Lemma [Lemma 10 in Ref. \cite{Maeda:2006pm}]- If $\Xi (x,r)$ is expanded as Eq. (\ref{expand})
with $N=1$ and $g_0(q) \neq q$, there exists an asymptotic solution satisfying
$x(r_s)=[g_1/(q-g_0)]^{1/(1-\alpha_1)}$ near $r=r_s$, and moreover it is the
unique solution to Eq. (\ref{nullgeo}) which is continuous at $r=r_s$.}

\vspace{3mm}

By virtue of this lemma, it is enough to investigate the features of the function $\Xi (x,r)$
for our current purpose. Hereafter we are interested in the future-directed outgoing null geodesics,
which correspond to positive $x(r_s)$, as can be seen from Eq. (\ref{nullgeodesiceq}).

From now on, we will address the existence of the null geodesics in odd- and even-dimensional cases separately.

\subsection{Odd dimensions}

As seen in the previous section, odd-dimensional spacetimes may have complicated
structures.

\subsubsection{Singularities at $r_s=0$}

We first analyze the geodesics from singularity at $r=0$ in odd dimensions.
In odd dimensions, the gravitational Eq. (\ref{basiceq2}) becomes
%
\begin{equation}
\sum_{m=1}^{(n+1)/2} c_m\left( \frac{\dot{v}}{v}\right)^{2m} =\frac{M(r)}{v^{n+1}}. \label{oddeom}
\end{equation}
%
Solving this with respect to $\dot{v}$, we have the following formal solution 
%
\begin{equation}
\dot{v} = V_{o}(v,r). \label{7vd}
\end{equation}
%
The formal integration of this with respect to $v$ is
%
\begin{equation}
t(v,r)=\int_1^v \frac{ds}{V_{o}(s,r)}.
\end{equation}
%

Let us assume the smooth distribution of matter near the center($r=0$) as
%
\begin{equation}
M(r)=M_0+M_2r^2+M_4r^4+\dots . \label{7massdistribc}
\end{equation}
%
Then $t(v,r)$ behaves near $r=0$ as
%
\begin{equation}
t(v,r)=t(v,0)+\frac{r^2}{2}\chi_2 (v)+\dots. \label{tvr}
\end{equation}
%
The orbit of the singularities is given by $t_s(r)=t(0,r)$. Since the definition
of $\chi_2 (v)$ is complicated, we will not write down its explicit form.
Regardless of the theory, it is shown that $\chi_2 (0)$ is non-negative under the condition of
$R^{\prime}>0$ and $\dot{R}<0$ in Ref. \cite{Maeda:2006pm}.
Here we further assume that $\chi_2 (0)$ is positive, that is,  $\chi_2 (0)$ cannot achieve zero (this corresponds to the assumption for the initial data).
Using the relation, which always holds,
%
\begin{equation}
0=\left( \frac{\partial t}{\partial v}\right)_{r}\left( \frac{\partial v}{\partial r}\right)_{t}+\left( \frac{\partial t}{\partial r}\right)_{v}, \label{trivial}
\end{equation}
%
we have
%
\begin{equation}
v^{\prime}=\left( -r\chi_2 (v)+\dots \right) \dot{v}. \label{7vpvdc}
\end{equation}
%

Now we set $q=3$ in Eq. (\ref{deftheta}) and then we can expand
$\Xi (x,r)$, with the help of Eqs. (\ref{oddeom}),
(\ref{7massdistribc}) and (\ref{7vpvdc}), as
%
\begin{align}
\Xi (x,r)&= \left( \frac{v+rv^{\prime}}{r^3}\right) \left( 1+r\dot{v}\right) \notag \\
         &=\frac{x+\chi_2 (v)\left( \frac{M_0}{c_{\frac{n+1}{2}} }\right)^{1/(n+1)} }{r}+\mathcal{O}(r^0).
\end{align}
%
From this expansion, the existence of solution to the null geodesic Eq. (\ref{nullgeo}) with
%
\begin{equation}
x(0)=\frac{\chi_2 (0)}{2} \left( \frac{M_0}{c_{\frac{n+1}{2}}} \right)^{1/(n+1)} >0
\end{equation}
%
is guaranteed according to the lemma. Since the positivity of $x(0)$ means the null geodesic is future-directed and outgoing null, this shows that the singularities are at least locally naked.

\subsubsection{Singularities at $0<r_s< r_{ah}$}

Next let us consider the geodesics from singularities at $0<r_s< r_{ah}$.
We consider the initial condition of
%
\begin{equation}
M(r)=\tilde{M}_0+\tilde{M}_1(r-r_s)+\tilde{M}_2(r-r_s)^2+\dots. \label{7massdistribf}
\end{equation}
%
In the previous case of $r_s=0$, we assumed smoothness at the center which implies
no kinklike distribution of matter. Therefore the next-to-leading term of expansion is
the order of $\mathcal{O}(r^2)$.
However, $r=r_s \neq 0$ is not the symmetric center and then the next-to-leading term of expansion
will be $\mathcal{O}(r-r_s)$ in general.

Near the singularities, then, $t(v,r)$ behaves as
%
\begin{equation}
t(v,r)=t(v,0)+(r-r_s)\tilde{\chi}_1 (v)+\dots. \label{7dt}
\end{equation}
%
Following Ref. \cite{Maeda:2006pm}, it is shown that $\tilde{\chi}_1 (0)$ is non-negative. Here
we simply assume  that $\tilde{\chi}_1 (0)$ cannot achieve zero. Using Eqs. (\ref{trivial})
and  (\ref{7dt}), we see
%
\begin{equation}
v^{\prime}=(-\tilde{\chi}_1 (v)+\dots )\dot{v} .\label{7vpvdf}
\end{equation}
%
Setting $q=1$ in Eq. (\ref{deftheta}), we can expand $\Xi (\tilde{x},r)$,
with the help of Eqs. (\ref{oddeom}), (\ref{7massdistribf}) and (\ref{7vpvdf}), as
\begin{widetext}
%
\begin{equation}
\Xi (\tilde{x},r)
         =\frac{r_s}{r-r_s}\tilde{\chi}_1 (v) \left( \frac{\tilde{M}_0}{c_{\frac{n+1}{2}} }\right)^{1/(n+1)}\left( 1-r_s\left( \frac{\tilde{M}_0}{c_{\frac{n+1}{2}} }\right)^{1/(n+1)}\right) +\mathcal{O}((r-r_s)^0) .
\end{equation}
%
\end{widetext}
According to Eqs. (\ref{rah}) and (\ref{7massdistribf}), $r_{ah}$ is
%
\begin{equation}
r_{ah}=\left( \frac{c_{\frac{n+1}{2}}}{\tilde{M}_0}\right)^{1/(n+1)}.
\end{equation}
%
From this expansion, the existence of a solution to the null
geodesic Eq. (\ref{nullgeo}) from singularities at $r=r_s$ ($0< r_s < r_{ah}$) with
\begin{widetext}
%
\begin{equation}
\tilde{x}(r_s)=r_s\tilde{\chi}_1 (0) \left( \frac{\tilde{M}_0}{c_{\frac{n+1}{2}} }\right)^{1/(n+1)}\left( 1-r_s\left( \frac{\tilde{M}_0}{c_{\frac{n+1}{2}} }\right)^{1/(n+1)}\right) >0
\end{equation}
%
\end{widetext}
is proved according to the lemma. This means that the singularities are at least locally naked.

\subsection{Even dimensions}

Finally we shall consider the even-dimensional spacetimes. In even dimensions
the gravitation field equation of (\ref{basiceq2}) becomes
%
\begin{equation}
\sum_{m=1}^{n/2} c_m\left( \frac{\dot{v}}{v}\right)^{2m} =\frac{M(r)}{v^{n+1}}. \label{eveneom}
\end{equation}
%
Then we obtain formally
%
\begin{equation}
\dot{v} = V_{e}(v,r). \label{8vd}
\end{equation}
%
The integration of this with respect to $v$ is
%
\begin{equation}
t(v,r)=\int_1^v \frac{ds}{V_{e}(s,r)}.
\end{equation}
%

We also consider the smooth distribution of matter as
%
\begin{equation}
M(r)=\bar{M}_0+\bar{M}_2r^2+\bar{M}_4r^4+\dots. \label{8massdistrib}
\end{equation}
%
Then $t(v,r)$ behaves like
%
\begin{equation}
t(v,r)=t(v,0)+\bar{\chi}_2 (v)\frac{r^2}{2}+\dots
\end{equation}
%
As in the case of odd dimensions, we assume that $\bar{\chi}_2 (0)$ is positive.
According to Eq. (\ref{trivial}), we have
%
\begin{equation}
v^{\prime}=\left( -r\bar{\chi}_2 (v)+\dots \right) \dot{v}. \label{8vpvdc}
\end{equation}
%
Setting $q=\frac{3n+1}{n+1}$ in Eq. (\ref{deftheta}), we can expand
$\Xi (\bar{x},r)$, with the help of Eqs. (\ref{eveneom}), (\ref{8massdistrib}) and (\ref{8vpvdc}),
as
%
\begin{align}
\Xi (\bar{x},r)
         =\frac{\bar{x}+\bar{\chi}_2 (v)\bar{x}^{-1/n}\left( \frac{\bar{M}_0}{c_{\frac{n}{2}} }\right)^{1/n}}{r}+\mathcal{O}(r^{-2/(n+1)}) .
\end{align}
%
From this expansion, the existence of solution to the null geodesic Eq. (\ref{nullgeo}) with
%
\begin{equation}
\bar{x}(0)=\left( \frac{n+1}{2n}\right)^{n/(n+1)} \bar{\chi}_2 (0)^{n/(n+1)} \left( \frac{\bar{M}_0}{c_{\frac{n}{2}}} \right)^{1/(n+1)} >0
\end{equation}
%
is guaranteed according to the lemma. Then the singularity is at least locally naked in this case too.

\section{Strength of singularity}

We finally compute the strength of naked singularity, which is the most
important feature characterizing singularities. We define
%
\begin{align}
{\cal R} =R_{\mu \nu}k^{\mu}k^{\nu},
\end{align}
%
where $k^{\mu}=\frac{dx^{\mu}}{d\lambda}$ is the tangential vector of future out-going
null geodesics from singularity and $\lambda$ is the affine parameter of the geodesics.
We will check the behavior of ${\cal R}$ near the singularity (See Ref. \cite{Abdolrahimi:2009dc}
 for a more systematic definition of the strength of singularity).

Consider the future outgoing radial null geodesics from singularities.
We have a relation from the metric of Eq. (\ref{metric2}) as follows
%
\begin{equation}
k^t=R^{\prime}k^r \label{tang}
\end{equation}
%
By a straightforward calculation using  Eq. (\ref{tang}), we have
%
\begin{equation}
{\cal R} =\frac{nF(r)^{\prime}}{\sum_{m=1}^{k}2mc_mR^{n-2m+2}\dot{R}^{2m-2}}R^{\prime}(k^r)^2.
\end{equation}
%
The radial null geodesic equation is
%
\begin{equation}
\frac{dk^r}{d\lambda}+\left( \frac{R^{\prime \prime}}{R^{\prime}}+2\dot{R}^{\prime}\right) (k^r)^2 =0.
\end{equation}
%

We need to consider odd- and even-dimensional cases separately.

\subsection{Odd dimensions}

\subsubsection{Singularity at $r=0$}

Near the singularity at $r=0$, we can compute ${\cal R}$ as
%
\begin{equation}
{\cal R} \sim 3n\left( \frac{M_0}{c_{\frac{n+1}{2}}}\right)^{2/(n+1)} (k^r)^2.
\end{equation}
%
The geodesic equation near the singularity becomes
%
\begin{equation}
\frac{dk^r}{d\lambda}+\frac{2}{r}(k^r)^2\simeq 0. \label{geooddc}
\end{equation}
%
The integration of Eq. (\ref{geooddc}) gives us
%
\begin{equation}
r \sim \lambda^{1/3}.
\end{equation}
%
We now evaluate the behavior of ${\cal R}$ near the singularities as
%
\begin{equation}
{\cal R} \sim \lambda^{-\frac{4}{3}}.
\end{equation}
%
In odd-dimensional cases, we confirmed that ${\cal R}$ 
diverges at $r=0$ and the behavior of divergence does not 
depend on the spacetime dimensions.

\subsubsection{Singularity at $0<r_s<r_{ah}$}

Near the singularities at $r=r_s$, ${\cal R}$ behaves as
%
\begin{equation}
{\cal R} \sim \frac{nr_s}{r-r_s}\left( \frac{\tilde{M}_0}{c_{\frac{n+1}{2}}}\right)^{2/(n+1)} \left( 1+\frac{r_s}{n+1}\frac{\tilde{M}_1}{\tilde{M_0}}\right) (k^r)^2 .
\end{equation}
%
The geodesic equation becomes
%
\begin{equation}
\frac{dk^r}{d\lambda}+C(k^r)^2\simeq 0, \label{geooddnc}
\end{equation}
%
where $C$ is a constant depending on the initial data.
From Eq. (\ref{geooddnc}) we see
that
\begin{equation}
r-r_s \sim \lambda ,
\end{equation}
%
near the singularity ($\lambda \simeq 0$). Then we can see
%
\begin{equation}
{\cal R} \sim \lambda^{-1}.
\end{equation}
%
As a result,  ${\cal R}$ diverges at $r=r_s$ ($0<r_s<r_{ah}$) and
the behavior of divergence does not depend on the spacetime dimensions.

\subsection{Even dimensions}

\subsubsection{Singularity at $r=0$}
We can evaluate ${\cal R}$ near the central singularity as
%
\begin{equation}
{\cal R} \sim (3n+1)\left( \frac{\bar{M}_0}{\bar{x}(0)c_{\frac{n}{2}}}\right)^{2/n}\frac{(k^r)^2}{r^{4/(n+1)}}.
\end{equation}
%
The geodesic equation near the singularity is
%
\begin{equation}
\frac{dk^r}{d\lambda}+\frac{2n}{(n+1)r}(k^r)^2\simeq 0 .\label{geoevenc}
\end{equation}
%
The integration of Eq. (\ref{geoevenc}) implies
%
\begin{equation}
r \sim \lambda^{\frac{n+1}{3n+1}}.
\end{equation}
%
Thus, near singularity, we have  
%
\begin{equation}
{\cal R} \sim \lambda^{-4\left( \frac{n+1}{3n+1}\right) }\,.
\end{equation}
%

In the even-dimensional cases, ${\cal R}$ diverges at $r=0$ and its behavior depends on the
dimensions. This feature is different from odd-dimensional cases.

\section{Summary and discussion}

In this paper we considered gravitational collapse of a spherical inhomogeneous dust cloud in
the Lovelock gravity with any dimension. We found that the formation
of apparent horizon depends on the dimensions, that is, if it is odd or even.
In even dimensions, noncentral singularities ($r \neq 0)$ will be safely wrapped by apparent horizons
and only ``central" singularities at $r=0$ could be naked.
In odd dimensions, on the other hand, even those singularities which form at $0\leq r \leq r_{ah}$ may
not be wrapped by apparent horizons and they could be naked. We also studied the
future-directed outgoing null geodesics from singularities. Then we showed
the existence of radial null geodesics from singularities in odd and even dimensions separately.
This is a clear  violation of cosmic censorship conjecture (at least the
strong version). As we mentioned in the introduction, singularities cannot be naked in more than six-dimensional
spacetime in the naive dimensional extension of the Einstein gravity. Therefore the Lovelock corrections worsen the situation
in the aspect of the CCC. These results coincide with those of the spherical collapse of inhomogeneous
dust in the Gauss-Bonnet theory, which is special case of the Lovelock theory, and the null dust
collapse in the Lovelock theory.

Compared with the Gauss-Bonnet gravity, we found that the Lovelock term changes the nature
of singularities more drastically. In the Gauss-Bonnet case, only massless naked singularities could form in more than six dimensions.
However, we found that massive naked singularities can form in all odd dimensions
in the Lovelock gravity.

We also examined the strength of the singularities and found that $R_{\mu\nu}k^\mu k^\nu$
, with the null tangent $k^\mu$ , diverges at the singularities for all cases  considered here. We also showed that the behavior of divergence depends on the spacetime dimensions in even dimension. However, it does not depend on the spacetime dimensions in the odd case. It is a characteristic feature of the Lovelock gravity.

In this paper we had several assumptions which should be relaxed in future studies.
First of all, we focused on the marginally bound cases with $W(r)=1$.
But, our analysis can be easily extended to the nonmarginally bound cases $(W(r)\neq 1)$.
We assumed that the matter is a dust cloud.
From the standpoint of the CCC, we should deal with
more generic matter fields. We also assumed here spherical symmetry.
Finally, most of our work is just the local analysis of singularity.
The global visibility of singularity is still unclear.
These are some of the remaining issues which will be hopefully addressed in the near future.

\begin{acknowledgments}
 We thank Tomohiro Takahashi, Kentaro Tanabe, Shuichiro Kinoshita,
Ryosuke Mizuno and Jiro Soda for their useful comments and discussions. SO thanks
Professor Takashi Nakamura for his continuous encouragement. SO are supported by the
Grant-in-Aid for the Global COE Program ``The Next Generation of Physics, Spun from Universality
and Emergence'' from the Ministry of Education, Culture, Sports, Science
and Technology (MEXT) of Japan.  TS is partially supported by
Grant-Aid for Scientific Research from Ministry of Education, Science,
Sports and Culture of Japan (Nos.~21244033,~21111006,~20540258 and
19GS0219), the Japan-U.K. Research Cooperative Programs. TS and SJ acknowledge support
under Indo-Japan (DST-JSPS) project.

\end{acknowledgments}

\appendix

\section{Junction between inner and outer solutions}

In this appendix, we will address the junction between the inner and outer solutions.
We suppose that the inner solution is
the collapsing dust cloud presented in the main text and the outer solution is a static vacuum black
hole. We follow the argument in Refs. \cite{Maeda:2006pm, Poisson:2004}, where they discussed
the similar junctions in the Gauss-Bonnet theory and Einstein theory.

Let $\Sigma $ to be the boundary of the two regions. The inner solution 
to Eqs. (\ref{metric2}) and
(\ref{A}) is
%
\begin{equation}
ds^2=-dt^2+R^{\prime}{}^2dr^2+R^2d\Omega_n^2,
\end{equation}
%
and the outside solution to Eq. (\ref{solution}) is
%
\begin{equation}
ds^2=-f(R)dT^2+\frac{1}{f(R)}dR^2+R^2d\Omega_n^2 .
\end{equation}
%
We suppose that $\Sigma$ is described by parametric equations $R=R_{\Sigma}(t)$
and $T=T_{\Sigma}(t)$. In addition, it is natural to think that the boundary is the comoving,
that is, $r=r_0=$constant because of dust.
The induced metric on $\Sigma$ from the metric of the inner region is written as
%
\begin{equation}
ds^2=-dt^2+R_{\Sigma}^2d\Omega_n^2 .
\end{equation}
%
On the other hand, using the metric of the outer region is also rewritten as
%
\begin{equation}
ds^2=-\left( f(R_{\Sigma})\dot{T}^2_{\Sigma}-\frac{\dot{R_{\Sigma}}^2}{f(R_{\Sigma})}\right) dt^2
+R_{\Sigma}^2d\Omega_n^2.
\end{equation}
%
Of course, they should be identical and then we see that
%
\begin{equation}
1=\left( f(R_{\Sigma})\dot{T}^2_{\Sigma}-\frac{\dot{R_{\Sigma}}^2}{f(R_{\Sigma})}\right)
\end{equation}
%
holds.

The extrinsic curvatures of $\Sigma$ evaluated from inner metric are
%
\begin{align}
{}^-K^t_t&=0 ,\\
{}^-K^i_j&=\delta^i_j R_{\Sigma}^{-1}.
\end{align}
%
In terms of the metric of the outer region, the extrinsic curvatures also have another expression as
%
\begin{align}
{}^{+}K^t_t&=\frac{\dot{\beta}}{\dot{R_{\Sigma}}}\left( f(R_{\Sigma})\dot{T}^2_{\Sigma}-\frac{\dot{R_{\Sigma}}^2}{f(R_{\Sigma})}\right)^{-1},\\
{}^{+}K^i_j&=\delta^i_j \frac{\beta}{R_{\Sigma}},
\end{align}
%
where
%
\begin{equation}
\beta =\sqrt{f+\dot{R}^2}=f\dot{T}.
\end{equation}
%
The continuity of the extrinsic curvature implies us
%
\begin{equation}
\beta =1.
\end{equation}
%
From the definition of $\beta$, we have the following two equations
%
\begin{align}
\frac{dT_{\Sigma}}{dt}&=f(R_{\Sigma})^{-1}, \\
\left( \frac{dR_{\Sigma}}{dt}\right)^2 &=1-f(R_{\Sigma}),
\end{align}
%
where
%
\begin{equation}
f(R_{\Sigma})=1-R^2_{\Sigma}\psi (R_{\Sigma}).
\end{equation}
%
Then we see that
%
\begin{equation}
\left( \frac{\dot{R}_{\Sigma}}{R_{\Sigma}}\right)^2 =\psi (R_{\Sigma})\label{newcond}
\end{equation}
%
is required.

It is remembered that $\psi$ is the solution to Eq. (\ref{defpsi}) as in the form of
%
\begin{equation}
\sum_{m=1}^k c_m \psi^{m} =\frac{\mu}{R^{n+1}_{\Sigma}},
\end{equation}
%
and $\dot{R}$ is the solution to Eq. (\ref{basiceq})
%
\begin{equation}
\sum_{m=1}^k c_m \left( \frac{\dot{R}_{\Sigma}}{R_{\Sigma}}\right)^{2m} =\frac{F(r_0)}{R^{n+1}_{\Sigma}} .
\end{equation}
%
As seen easily, Eq. (\ref{newcond}) is satisfied only if
%
\begin{equation}
\mu =F(r_0).
\end{equation}
%
This shows that the inner region can be naturally joined to the outer region if $\mu =F(r_0)$ is satisfied.

\section{Homogeneous dust collapse}

In this section, we briefly discuss homogeneous case which is just a special case 
of inhomogeneous collapse. In this case, the spacetime metric becomes 
%
\begin{equation}
ds^2=-dt^2+a^2(t) \big( dr^2+r^2d\Omega_n^2 \big) .
\end{equation}
%
Equation (\ref{basiceq}) reduces to 
%
\begin{equation}
\sum_{m=1}^kc_m\left( \frac{\dot{a}}{a}\right)^{2m} =\frac{M}{a^{n+1}},
\end{equation}
%
where $M$ is a constant. From this equation, we can see that the behavior 
of $a$ near the singularity ($a=0$) as
%
\begin{align}
a(t) \sim & (t_s-t) \ \ &\text{(in odd dimensions)}\\
a(t) \sim & (t_s-t)^{n/n+1} \ \ &\text{(in even dimensions)}
\end{align}
%
where $t_s$ stands for the epoch of the singularity formation.
From these behaviors, it turns out that singularity 
is ingoing-null in odd dimensions and is spacelike 
in even dimensions according to Ref. \cite{senovilla}. 
In the Gauss-Bonnet homogeneous case \cite{Maeda:2006pm}, 
singularity in five dimension is ingoing-null and is 
spacelike in dimensions higher than six. 
Therefore, as in the main text, we can confirm that the Lovelock terms changes 
the nature the singularity. 


\end{document}